**Quantum computing with defects**

*J. R. Weber[1], W. F. Koehl[1], J. B. Varley[1], A. Janotti, B. B. Buckley, C. G. Van de Walle, and D. D. Awschalom[2]*

*Center for Spintronics and Quantum Computation, University of California, Santa Barbara, CA 93106*

*[1] J.R.W., W.F.K., and J.B.V. contributed equally to this work.*

*[2] To whom correspondence should be addressed. E-mail: awsch@physics.ucsb.edu*

**Identifying and designing physical systems for use as qubits, the basic units of quantum information, are critical steps in the development of a quantum computer. Among the possibilities in the solid state, a defect in diamond known as the nitrogen-vacancy ($NV^{-1}$) center stands out for its robustness – its quantum state can be initialized, manipulated, and measured with high fidelity at room temperature. Here we describe how to systematically identify other deep center defects with similar quantum-mechanical properties. We present a list of physical criteria that these centers and their hosts should meet and explain how these requirements can be used in conjunction with electronic structure theory to intelligently sort through candidate defect systems. To illustrate these points in detail, we compare electronic structure calculations of the $NV^{-1}$ center in diamond with those of several deep centers in 4H silicon carbide (SiC). We then discuss the proposed criteria for similar defects in other tetrahedrally-coordinated semiconductors.**

A quantum computer is a device that would exploit the rules of quantum mechanics to solve certain computational problems more efficiently than allowed by Boolean logic (1). Over the past two decades, qubits have been implemented in a wide variety of materials, including atoms (2), liquids (3), and solids such as superconductors (4), semiconductors (5), and ion-doped insulators (6). Recently, the diamond $NV^{-1}$ has emerged as a leading qubit candidate because it is an individually-addressable quantum system that may be initialized, manipulated, and measured with high fidelity at room temperature (7). Interestingly, even though these successes stem largely from the defect's nature as a deep center (a point defect with highly-localized electronic bound states confined to a region on the scale of a single lattice constant), no systematic effort has been made to identify other deep centers that might behave similarly. We outline the physical features that such deep centers and their hosts should exhibit, and show how these criteria can be used to identify potential qubit candidates within a large class of defects structurally analogous to the diamond $NV^{-1}$. To aid in the illustration of these points, we compare density functional theory (DFT) calculations of the diamond $NV^{-1}$ with those of several defects found in 4H-SiC.

Searching for deep centers that behave like the diamond $NV^{-1}$ is worthwhile for several reasons. From an engineering perspective, it is currently quite difficult to grow and fabricate devices from diamond. The discovery of a similar defect in a more technologically mature host material might allow for more sophisticated implementations of single- and multi-qubit devices. Additionally, because deep centers and semiconductors as a whole exhibit a diverse set of physical characteristics, new areas of device functionality may potentially arise once the quantum properties of these defect systems are more fully explored. From a physics perspective, other deep centers with highly-controllable quantum states might help to resolve outstanding questions regarding the structure and dynamical properties of the diamond $NV^{-1}$, or of deep centers in general.

**Defect and Host Criteria for NV-like Systems**

Structurally, the diamond $NV^{-1}$ consists of a carbon vacancy and an adjacent substitutional nitrogen impurity. The bound states of this deep center are multi-particle states composed of six electrons: five contributed by the four atoms surrounding the vacancy, and one captured from the bulk. As shown in Fig. 1, the lowest energy bound state is a spin triplet ($^3A_2$) whose spin sublevels differ slightly in energy. The $m_s = 0$ and -1 sublevels of this ground state can be chosen to function as the qubit state, and coherent rotations between the two sublevels may be induced by applying microwave radiation tuned to the energy splitting between them. A spin-conserving optical transition exists between the $^3A_2$ state and an excited state triplet ($^3E$) 1.945 eV higher in energy. In addition, there exists a spin-selective decay path between these two states that includes a non-radiative transition from $^3E$ to an intermediate spin singlet ($^1A_1$). In combination, these transitions allow the center to be optically initialized and measured. That is, they allow the defect to be optically pumped into the $m_s = 0$ sublevel of $^3A_2$, and they cause the fluorescence intensity between $^3E$ and $^3A_2$ to be spin-dependent (8).

Two features of the diamond $NV^{-1}$ help to distinguish it from other solid state qubit systems. First, the center's highly localized bound states are well isolated from sources of decoherence. At room temperature, the ground state can exhibit extremely long spin coherence times of up to 1.8 ms (9). This is close to the regime needed for quantum error correction, given that manipulation rates greater than 200 MHz have been demonstrated (10, 11). Second, the structure of the defect's excited state manifold allows the defect to be optically initialized and measured with high fidelity under ambient conditions. Many other solid state systems are initialized via thermal equilibration, and therefore require cryogenic operating temperatures (12-14). And while other systems can be initialized optically (15), or can operate at room temperature (16), they currently can only be measured with high fidelity in an ensemble.

To reproduce these two features, there are several criteria that a candidate deep center and its host should meet. Specifically, centers should exhibit the following five characteristics (for simplicity, we restrict discussion of these characteristics to centers for which, like the diamond NV$^{-1}$, spin can be treated as a good quantum number):

D1) *A bound state that is suitable for use as a qubit.* This state must be paramagnetic and long-lived, and an energy splitting must exist between at least two of the state's spin sublevels. If the qubit state is to be manipulated via electron spin resonance, the size of this energy splitting must fall within an appropriate range of the radio frequency spectrum.

D2) *An optical pumping cycle that polarizes the qubit state.* This cycle will most likely consist of an optical transition from the ground state to an excited state, followed by a spin-selective decay path that includes one or more non-radiative transitions between states of differing spin multiplicity.

D3) *Luminescence to or from the qubit state that varies by qubit sublevel in some differentiable way, whether by intensity, wavelength, or other property.* If fluorescence from an excited state is used to probe the qubit, the fluorescent transition should be spin-conserving. In addition, the strength of this fluorescent transition, which depends on the lifetime of the excited state, should be large enough to enable efficient, high fidelity measurement of individual defect qubit states.

D4) *Optical transitions that do not introduce interference from the electronic states of the host.* All optical transitions used to prepare and measure the qubit state must be lower in energy than the energy required to transfer an electron into (out of) the center from (to) the electronic states of the host.

D5) *Bound states that are separated from each other by energies large enough to avoid thermal excitation between them.* If the energy difference between two bound states is too small, thermal excitations may couple states and destroy spin information.

In addition, an ideal crystalline host will have the following qualities, the final three of which should help to reduce decoherence in the defect:

H1) *A wide band gap*, so that it can accommodate a deep center that will satisfy requirement D4 above.

H2) *Small spin-orbit coupling,* in order to avoid unwanted spin-flips in the defect bound states.

H3) *Availability as high-quality, bulk or thin-film single crystals,* in order to avoid imperfections or paramagnetic impurities that could affect the deep center's spin state.

H4) *Constituent elements with naturally occurring isotopes of zero nuclear spin,* so that spin bath effects may be eliminated from the host via isotopic engineering (9).

It is relatively easy to identify hosts that satisfy criteria H1-H4, and this is discussed in detail in Section S1 of the Supporting Information. However, it is not as simple to predict whether a given defect will satisfy criteria D1-D5. Nevertheless, certain aspects of a defect's electronic structure can be predicted in a straightforward manner using first-principles calculations. For example, the spin of a defect bound state can be calculated, determining whether the defect is paramagnetic or not, which is a major component of criterion D1. However, it is difficult to accurately compute the energy splittings between the spin sublevels of a bound state. In the case of criterion D2, the defect-induced gap levels obtained via first-principles calculations can be used to predict whether a paramagnetic defect will possess internal optical transitions. In addition, the energies of these optical transitions can be calculated using constrained density functional theory, and excitonic effects can be included through the Bethe-Salpeter formalism (17). An explicit characterization of non-radiative decay paths, on the other hand, is much more challenging and is generally beyond the reach of current first-principles methods (18). The application of first-principles calculations to criterion D3 depends on the desired method of qubit measurement. In cases like the diamond NV$^{-1}$, where the *intensity* of luminescence to the qubit state should vary by spin sublevel, first-principles predictions are again limited by the challenging nature of non-radiative transitions. In cases where the *wavelength* of luminescence is meant to vary by spin sublevel, the small magnitude of the spin sublevel splittings (in the microwave range for qubits manipulated using electron spin resonance) renders it difficult to *quantitatively* evaluate luminescence energies based on first-principles calculations. Still, in combination with perturbation theory, first-principles calculations can provide information about the ordering of sublevels, and this can help guide experimental identification of observed optical transitions. Additionally, the excited-state lifetime discussed in criterion D3 can be characterized by calculating the magnitudes of the dipole matrix elements associated with the internal optical transitions. Finally, the properties associated with criteria D4 and D5 can be studied explicitly by analyzing the energies of calculated defect-induced gap levels, both relative to each other, and relative to the band edges of the host material.

Even though some aspects of D1-D5 are difficult to evaluate using standard first-principles methods, potential qubit candidates can still be identified by observing that a close relationship exists between the atomic configuration of a defect and its electronic structure (19). In tetrahedrally-coordinated semiconductors, vacancies and vacancy-related complexes similar to the NV$^{-1}$ center in diamond are likely to possess bound states with comparable physical properties. Within this group of defects, one can then use first-principles calculations to determine which members are compatible with criteria D1-D5. In the next section, we demonstrate in detail how this assessment can be made, using as an example the results of calculations performed for several defects in diamond and SiC.

**Formation Energies, Defect-Level Diagrams, and Configuration-Coordinate Diagrams**

Density functional theory calculations have become an indispensable tool for studying the properties of defects. Recent advances using hybrid functionals, which incorporate some degree of the Hartree-Fock exchange interaction, have led to very accurate descriptions of defect states by overcoming the well-known band-gap problem of traditional DFT. We apply this methodology to defect systems analogous to the $NV^{-1}$ defect in diamond. Specifically, we discuss defect formation energies, configuration-coordinate diagrams for defect excitations, as well as the arrangement of the defect-induced gap levels, which we discuss in terms of "defect-level diagrams" (DLDs).

One of the most important quantities that can be extracted from first-principles calculations is the formation energy ($E^f$) of a defect. $E^f$ provides information on the overall stability of a given defect, as well as the relative stabilities between different atomic configurations and charge states. $E^f$ determines the defect concentration through a Boltzmann relation (20):

$$C = N_S e^{-E^f / k_B T},$$

where $N_S$ is the number of possible defect sites. Strictly speaking, this expression is only valid in equilibrium; however, formation energies are informative even when defects are created in non-equilibrium processes, such as ion implantation. Specifically, the magnitude of $E^f$ still provides an indicator of which defects are most likely to form. Once a defect is formed, the relative stability of different charge states for a given defect is always determined by the dependence of $E^f$ on Fermi level, whatever the creation process of the defect.

**The NV Center in Diamond.** As explained in the *Methods* section, $E^f$ depends on the charge state of the defect and on the Fermi level ($\varepsilon_F$), which is referenced to the valence-band maximum (VBM) of the bulk host material. Figure 2 shows $E^f$ for the NV center in diamond, as well as for the carbon vacancy ($V_C$), and substitutional nitrogen ($N_C$). For each defect, only the charge state with the lowest $E^f$ is included at each value of $\varepsilon_F$. For a given $\varepsilon_F$, the charge state of a defect is equal to the slope of that defect's $E^f$ curve at that point. The kinks in the $E^f$ curves correspond to charge-state-transition levels, i.e., $\varepsilon_F$ values where the charge state of the defect changes. Figure 2 thus shows the range of $\varepsilon_F$ for which each charge state of the defect is stable; both the isolated vacancy and the NV center can be stable in the +1, 0, -1, or -2 charge states. Determining which charge states are stable, and under what conditions, is crucial to evaluating whether these vacancy-related defects satisfy criterion D1. This is because each charge state will correspond to a different spin configuration, with some having paramagnetic ground states and others not.

The ground state spin of each charge state can be determined by considering the defect's electronic structure. The electronic structure of vacancy-related centers in tetrahedrally-coordinated semiconductors can be understood in terms of atomic $sp^3$ orbitals and the corresponding single-particle levels. In an environment with tetrahedral symmetry, the four degenerate $sp^3$ dangling-bond (DB) orbitals neighboring a vacancy are split into a low-energy symmetric $a_1$ level and three degenerate $t_2$ levels (as seen in Fig. 3A for the -2 charge state of $V_C$, which is stable in N-doped diamond). Because of the high symmetry of the isolated vacancy, this level structure does not lead to a ground-state triplet. This can be achieved by placing an impurity atom next to the vacancy, thus shifting the $a_1$ level (becoming $a_1(1)$) and splitting the degeneracy of the $t_2$ levels into $a_1(2)$, $e_x$, and $e_y$ levels (21); Fig. 3B shows the DLD for $NV^{-1}$.

The $NV^{-1}$ defect is stable for $\varepsilon_F$ between 2.78 and 5.14 eV, consistent with the likely position of $\varepsilon_F$ in N-doped diamond (N being a deep donor located ~1.9 eV below the conduction band minimum (CBM) (22)). The associated single-particle eigenvalues are listed in Table 1; their occupation determines the spin state (i.e., a spin-one triplet). The location of the defect-induced gap levels illustrates the defect's compliance with D4 and D5, since they are well isolated from the bulk bands, with a relatively large spacing between occupied and unoccupied levels. As shown by the green dashed arrow in Fig. 3B, we can remove an electron from the spin-minority channel of the $a_1(2)$ level, and place it into one of the $e_i$ spin-minority levels, keeping the ground-state atomic configuration fixed. The corresponding absorption energy of 2.27 eV is shown in the configuration-coordinate diagram of Fig. 4A. If we subsequently allow the atomic positions to relax, maintaining the excited-state triplet electronic configuration, we obtain a zero-phonon line (ZPL) energy of 2.02 eV and a Frank-Condon shift of 0.26 eV, both in good agreement with experiment (23) as well as with recent calculations (24). This illustrates how the computationally accessible properties relevant to criterion D2 can be obtained.

**Defect Centers in SiC.** SiC shares many similarities with diamond, and recent experimental evidence indicates that it may also harbor deep centers suitable for quantum computing (25-28). We focus on the 4H polytype because large single crystals are readily available, and because its band gap (3.27 eV) is larger than that of 3C-SiC (2.39 eV) and 6H-SiC (3.02 eV) (29). Figure 2B shows $E^f$ for the silicon and carbon vacancies ($V_{Si}$ and $V_C$), as well as for substitutional nitrogen ($N_{Si}$ and $N_C$). $E^f$ is more than 4 eV larger for $N_{Si}$ than for $N_C$, so N has an extremely strong energetic preference to sit on a C site. This implies that only nitrogen-vacancy centers composed of a $N_C$ and a $V_{Si}$ will form in SiC. According to Fig. 2B, $N_C V_{Si}$ is stable in the 0, -1, and -2 charge states. Similar to the diamond $NV^{-1}$ defect, six electrons are confined to the $N_C V_{Si}^{-1}$ defect, which is stable for $\varepsilon_F$ between 1.60 and 2.83 eV. The levels for the corresponding DLD are listed in Table 1. Note that the degeneracy of the $e_i$ levels is lifted due to the lower symmetry of the crystal structure.

The calculated configuration-coordinate diagram for $N_C V_{Si}^{-1}$ is shown in Fig. 4B. Comparing these numbers with the diamond $NV^{-1}$ (Fig. 4A), we see that the vertical transitions are about half as large in the $N_C V_{Si}^{-1}$ center, while the relaxation energies are more than 75% smaller. The difference in transition energies can be attributed to the larger lattice constant of SiC compared with diamond: although the vacancy is surrounded by C atoms in both materials, the larger lattice constant of SiC leads to a smaller overlap among the $sp^3$ DB orbitals and therefore to a smaller splitting between $a_1(2)$ and $e_i$ levels.

It is interesting to also consider isolated vacancies in SiC. $V_{Si}^{-2}$ in SiC can support a spin-conserving triplet excitation because the broken tetrahedral symmetry of the host splits the $t_2$ levels (Figs. 3C and 3D). Our calculated formation energies in Fig. 2B show that the 0, -1, and -2 charge states are all stable. Similar to the diamond $NV^{-1}$ or SiC $N_CV_{Si}^{-1}$ defects, six electrons are bound to $V_{Si}^{-2}$, which is stable in *n*-type material for Fermi levels within 0.3 eV of the CBM. The DLD in Fig. 3*C* (energies in Table 1) shows that the spin-minority $a_1(2)$ level lies above the spin-majority $e_i$ levels, which raises the possibility that $V_{Si}$ may also exhibit a ground-state triplet when occupied with four electrons instead of six. This is indeed borne out by explicit calculations, as shown in Fig. 3*D*. In principle, a similar situation could occur by removing two electrons from the $N_CV_{Si}^{-1}$ defect; however, $N_CV_{Si}^{+1}$ is not a stable charge state, as evident from Fig. 2*B*. However, the close proximity of the spin-majority $a_1(2)$ level to the VBM is cause for concern in light of criterion D4. Finally, we note that $V_{Si}^{-1}$ forms a ground state quartet (i.e., spin 3/2, see Table 1), and therefore this defect does not allow for spin-conserving triplet excitations.

**Discovering NV Analogs in Other Material Systems**

Moving beyond SiC, it is important to establish some general guidelines and procedures for identifying defects that may be analogous to the NV center in diamond. For the purposes of this discussion, we will focus on tetrahedrally-coordinated compound semiconductors, considering both cation and anion vacancies.

**Cation Vacancies.** In cation vacancies, the defect levels are determined by interacting anion $sp^3$ DBs. Since anion DBs lie close to the VBM (30) (Fig. 5*B*), the $t_2$ vacancy levels will tend to be located in the lower half of the band gap. To satisfy criterion D4, these $t_2$ levels should be well separated from the VBM. The $t_2$ levels will be split by (1) Jahn-Teller distortions, (2) the presence of an impurity, and/or (3) crystal-field splitting in hosts with lower than cubic symmetry. This splitting should be sufficiently large to satisfy criterion D5, but small enough to satisfy D4, and to avoid pushing the $a_1(2)$ level too close to the VBM. The energy position of the anion DB orbitals and the splitting between the $a_1$ and $t_2$ vacancy orbitals ($\Delta_{CV}$ - Fig. 5*B*) are therefore important quantities for identifying new defect systems for quantum computing applications, and in Section S2 of the Supporting Information we address how the choice of host and defect center impacts their value.

Further design flexibility is added by placing an impurity next to the vacancy, thus creating a complex. The energy of the impurity's $sp^3$ orbital relative to that of the host anion DB orbital affects the splitting between the $a_1(2)$ and $e_i$ orbitals ($\delta_{CV}$ – Fig. 5C). If the impurity DB orbital is significantly lower in energy than the host anion DB, the splitting $\delta_{CV}$ will be large. Too large a $\delta_{CV}$ value (relative to the band gap) is undesirable, since it might push the $a_1(2)$ level close to or below the VBM. An attractive interaction is needed between the vacancy and impurity in order for the complex to form. Therefore, since cation vacancies tend to be negatively charged, we should choose impurities that act as donors, i.e., elements to the right of the host anion in the periodic table. An example of such a defect is the SA-center in ZnSe, which is a complex formed by a Zn vacancy and a donor impurity. In the positive charge state (which would be stable in $p$-type material), this defect gives rise to a ground-state triplet with six electrons (see Section S3 of the Supporting Information). It remains to be determined whether this defect satisfies all the other proposed criteria.

**Anion Vacancies.** Anion vacancies are less likely to lead to triplet configurations, since the cation DBs that give rise to their defect levels tend to be located in the upper part of the band gap (30) (Fig. 5B) and the vacancy orbital splitting ($\Delta_{AV}$) will tend to push the $t_2$ orbitals close to or above the conduction-band minimum. This is indeed what happens for an oxygen vacancy in ZnO, for which only the $a_1(1)$ level lies within the band gap (31). To avoid this, the semiconductor needs to have a large enough band gap, the cation $sp^3$ DB orbitals need to be well below the CBM, and the vacancy orbital splitting ($\Delta_{AV}$) needs to be small. These criteria are met in AlN, in which the $V_N$ has $t_2$ levels within the band gap. The arguments about further splitting of the levels are similar to our discussion of cation vacancies. Regarding the choice of impurity, since anion vacancies tend to act as donors, one might think that acceptor-type impurities might be the best choice, in order to maximize attraction. However, electron counting then reveals that a level occupation similar to that of the diamond $NV^{-1}$ cannot be achieved because this requires that the anion vacancy (or complex) be in a negative charge state. But in AlN, $V_N^{-1}$ is stable if $\varepsilon_F$ is in the upper part of the band gap, and a *donor* impurity will fulfill the requirements of 1) being attracted to the vacancy and 2) supplying additional electrons to achieve the desired orbital occupation.

**Beyond NV Analogs**

The world of deep centers is vast, and only one small subset has been discussed here in detail. Future work is needed to determine which other classes of deep centers are compatible with the defect and host criteria that have been outlined. For example, many isolated substitutional or interstitial impurities act as deep centers (19), but no such center satisfying D1-D5 has been identified so far. In the octahedrally-coordinated hosts MgO and CaO, optical spin polarization has been reported in vacancy-related complexes with $D_{4h}$ symmetry (32, 33), but more exploration is required to determine what other features these centers have in common with the $NV^{-1}$ center in diamond. Still other classes of deep centers become open to investigation if the stipulation that spin be a good quantum number is removed. In this case, optical selection rules are relaxed, and alternative mechanisms of optical polarization may then be possible (34).

**Methods**

The following expression gives $E^f$ for the NV center in diamond in charge state $q$:

$$E^f\left[C:NV^q\right] = E_{tot}\left[C:NV^q\right] - E_{tot}\left[C:bulk\right] - \mu_N + \mu_C + q\left(\varepsilon_F + \varepsilon_{VBM}^{bulk} + \Delta V\right)$$

The $E_{tot}[C:X]$ terms are the total energies of the diamond supercell with the $NV^q$ defect and of the bulk supercell. The $\mu_X$ terms are the chemical potential references used for N and C. For diamond, $\mu_C$ is simply the energy per C atom in the crystal. $\varepsilon_F$ is the Fermi level, referenced to the valence-band maximum (VBM) in the bulk, $\varepsilon_{VBM}^{bulk}$; the $\Delta V$ term is used to align the bulk VBM to that of the defect supercell (20, 35).

The first-principles calculations were performed using supercells of 64 atoms for C in the diamond structure, and 96 atoms for SiC in the 4H-polytype ($C^4_{6v}$ space group), with finite-size corrections for the charged-defect calculations (35). Projector augmented wave pseudopotentials were used as implemented in the Vienna *ab initio* Simulation Package (VASP) (36, 37). We used a 400 eV plane-wave cutoff, and a 2×2×2 special-*k*-point mesh to carry out integration over the Brillouin zone. The hybrid functional calculations (including atomic relaxations) were performed within the HSE06 formalism (38, 39). The calculated band gaps are 5.36 eV for diamond and 3.17 eV for 4H-SiC. Defects in 4H-SiC can occur on two possible inequivalent Si or C sites; tests indicate the corresponding energies differ by less than 0.1 eV. The results reported here are for the hexagonal site. For the NV defect in SiC, there are two choices for the position of the substitutional N atom: one associated with the single longer bulk Si-C bond length (along c-axis), and the other associated with the three shorter bulk Si-C bond lengths. For our calculations, we chose the site associated with the shorter Si-C bond. All defect excitation energies were calculated using constrained DFT, by removing an electron out of an occupied defect level, and placing it into an occupied defect level. We note that transitions between internal defect levels are likely to be more accurately calculated than defect-to-band transitions (40). For the purposes of assessing our criteria, this trend is advantageous, since information about defect-to-band transitions is used only qualitatively in determining whether such transitions are suppressed.

**Acknowledgements**

This work was supported by ARO, AFOSR, and by the NSF MRSEC Program. It made use of the CNSI Computing Facility and TeraGrid (TACC and SDSC) with support from the NSF. We are grateful to G. D. Fuchs, F. J. Heremans, and D. M. Toyli for useful discussions.

**Competing Financial Interests statement**

The authors declare that they have no competing financial interests.

**Author Contributions**

All authors performed research and wrote the paper.

**References**

1)  Childs AM, van Dam W (2010) Quantum algorithms for algebraic problems. *Rev. Mod. Phys.* 82:1-52.

2)  Monroe C (2002) Quantum information processing with atoms and photons. *Nature* 416:238-246.

3)  Cory DG, et al. (2000) NMR based quantum information processing: Achievements and prospects. *Fortschr. Phys.* 48:875-907.

4)  Devoret MH, Martinis JM (2004) Implementing qubits with superconducting integrated circuits. *Quantum Inform. Process.* 3:163-203.

5)  Hanson R, Awschalom DD (2008) Coherent manipulation of single spins in semiconductors. *Nature* 453:1043-1049.

6)  Bertaina S, et al. (2007) Rare-earth solid-state qubits. *Nature Nano.* 2:39-42.

7)  Neumann P, et al. (2008) Multipartite entanglement among single spins in diamond. *Science* 320:1326-1329.

8)  Manson NB, Harrison JP, Sellars MJ (2006) Nitrogen-vacancy center in diamond: Model of the electronic structure and associated dynamics. *Phys. Rev. B* 74:104303.

9)  Balasubramanian G, et al. (2009) Ultralong spin coherence time in isotopically engineered diamond. *Nature Mater.* 8:383-387.

10) DiVincenzo DP (2000) The physical implementation of quantum computation. *Fortschr. Phys.* 48:771-783.

11) Fuchs GD, Dobrovitski VV, Toyli DM, Heremans FJ, Awschalom DD (2009) Gigahertz dynamics of a strongly driven single quantum spin. *Science* 326:1520-1522; published online 19 November 2009 (10.1126/science.1181193).

12) Kane BE (1998) A silicon-based nuclear spin quantum computer. *Nature* 393:133–137.

13) Petta JR, et al. (2005) Coherent manipulation of coupled electron spins in semiconductor quantum dots. *Science* 309:2180-2184.

14) Berezovsky J, Mikkelsen MH, Stoltz NG, Coldren LA, Awschalom DD (2008) Picosecond coherent optical manipulation of a single electron spin in a quantum dot. *Science* 320:349-352.

15) Wesenberg JH, Mølmer K, Rippe L, Kröll S (2007) Scalable designs for quantum computing with rare-earth-ion-doped crystals. *Phys. Rev. A* 75:012304.

16) Nellutla S, et al. (2007) Coherent manipulation of electron spins up to ambient temperatures in $Cr^{5+}(S=1/2)$ doped $K_3NbO_8$. *Phys. Rev. Lett.* 99:137601.

17) Ma Y, Rohlfing M (2007) Quasiparticle band structure and optical spectrum of $CaF_2$. *Phys. Rev. B* 75:205114.

18) Gali A, Fyta M, Kaxiras E (2008) *Ab initio* supercell calculations on nitrogen-vacancy center in diamond: Electronic structure and hyperfine tensors. *Phys. Rev. B* 77:155206.

19) Lannoo M, Bourgoin J (1981) *Point Defects in Semiconductors I: Theoretical Aspects*, (Springer-Verlag, New York), pp 9-22, 68-92.

20) Van de Walle CG, Neugebauer J (2004) First-principles calculations for defects and impurities: Applications to III-nitrides. *J. Appl. Phys.* 95:3851-3879.

21) Lenef A, Rand SC (1996) Electronic structure of the N-V center in diamond: Theory. *Phys. Rev. B* 53:13441-13455.

22) Heremans FJ, Fuchs GD, Wang CF, Hanson R, Awschalom DD (2009) Generation and transport of photoexcited electrons in single-crystal diamond. *Appl. Phys. Lett.* 94:152102.

23) Davies G, Hamer MF (1976) Optical studies of the 1.945 eV vibronic band in diamond. *Proc. R. Soc. London, Ser. A* 348:285-298.

24) Gali A, Janzén E, Deak P, Kresse G, Kaxiras E (2009) Theory of spin-conserving excitation of the N-V$^-$ center in diamond. *Phys. Rev. Lett.* 103:186404.

25) Son NT, Zolnai Z, Janzén E (2003) Silicon vacancy related T$_{V2a}$ center in 4H-SiC. *Phys. Rev. B* 68:205211.

26) Mizuochi N, et al. (2002) Continuous-wave and pulsed EPR study of the negatively charged silicon vacancy with S=3/2 and $C_{3v}$ symmetry in *n*-type 4*H*-SiC. *Phys. Rev. B* 66:235202.

27) Orlinski SB, Schmidt J, Mokhov EN, Baranov PG (2003) Silicon and carbon vacancies in neutron-irradiated SiC: A high-field electron paramagnetic resonance study. *Phys. Rev. B* 67:125207.

28) Baranov PG, et al. (2007) Spin polarization induced by optical and microwave resonance radiation in a Si vacancy in SiC: A promising subject for the spectroscopy of single defects. *JETP Lett.* 86:202-206.

29) Madelung O (2004) *Semiconductors: Data Handbook*, 3rd edition (Springer-Verlag, New York).

30) Harrison W (1989) *Electronic structure and the properties of solids*, (Dover Publications, Inc., Mineola, NY).

31) Janotti A, Van de Walle CG (2005) Oxygen vacancies in ZnO. *Appl. Phys. Lett.* 87:122102.

32) Tanimoto DH, Ziniker WM, Kemp JO (1965) Optical Spin Polarization in M-like centers in CaO. *Phys. Rev. Lett.* 14:645-647.

33) Henderson B (1976) Optical pumping cycle of exchange-coupled $F^+$-centre pairs in MgO and CaO. *J. Phys. C: Solid State Phys.* 9:2185-2195.

34) Geschwind S (1972) in *Electron Paramagnetic Resonance*, ed Geswchind S (Plenum Press, New York), pp 372-386.

35) Freysoldt C, Neugebauer J, Van de Walle CG (2009) Fully ab initio finite-size corrections for charged supercell calculations. *Phys. Rev. Lett.* 102:016402.

36) Kresse G, Furthmuller J (1996) Efficient iterative schemes for *ab initio* total-energy calculations using a plane-wave basis set. *Phys. Rev. B* 54:11169.

37) Kresse G, Joubert J (1999) From ultrasoft pseudopotentials to the projector augmented-wave method. *Phys. Rev. B* 59:1758-1775.

38) Heyd J, Scuseria GE, Ernzerhof M (2003) Hybrid functionals based on a screened Coulomb potential. *J. Chem. Phys.* 118**:**8207-8215.

39) Marsman M, Paier J, Stroppa A, Kresse G (2008) Hybrid functionals applied to extended systems. *J. Phys. Condens. Mater.* 20**:**064201.

40) Alkauskas A, Broqvist P, and Pasquarello A (2008) Defect energy levels in density functional calculations: Alignment and band gap problem. *Phys. Rev. Lett.* 101: 046405.

**Figure Legends**

Figure 1.   Multiplet structure of the NV$^{-1}$ center in diamond.  A 1.945 eV spin-conserving optical transition exists between the ground ($^3$A$_2$) and excited ($^3$E) state triplets.  Transitions from the $m_s = \pm 1$ sublevels of $^3$E to an intermediate spin singlet ($^1$A$_1$) are much stronger than those from the $m_s = 0$ sublevel.  The spin-selective nature of this decay path can be used in conjunction with the 1.945 eV transition to optically polarize and measure the spin state of $^3$A$_2$.

Figure 2.   Formation energy, $E^f$, as a function of Fermi level, $\varepsilon_F$.  $E^f$ was calculated for various defects in (*A*) diamond and (*B*) 4H-SiC (in C-rich conditions).  The shaded areas show the range of stability of NV$^{-1}$ in diamond, and V$_{Si}^0$ (blue), N$_C$V$_{Si}^{-1}$ (green) and V$_{Si}^{-2}$ (purple) in SiC.

Figure 3.   Defect-level diagrams for vacancy-related complexes.  These diagrams show the single-particle defect states for (*A*) the V$_C^{-2}$ and (*B*) the NV$^{-1}$ in diamond, as well as for (*C*) the V$_{Si}^{-2}$ and (*D*) the V$_{Si}^0$ in 4H-SiC.  The spin-majority (spin-minority) channel is denoted by upward (downward) pointing arrows.

Figure 4.   Configuration-coordinate diagrams for spin-conserving triplet excitation.  Excitation cycles for (*A*) the NV$^{-1}$ center in diamond and (*B*) the N$_C$V$_{Si}^{-1}$ center in SiC are shown.  Absorption, emission, and zero-phonon line (ZPL) transitions are indicated, along with their energies.

Figure 5.   Development of defect level structure in tetrahedrally coordinated compound semiconductors.  Atomic $sp^3$ dangling bonds (*A*) interact to form $a_1$ and $t_2$ levels in an ideal vacancy (*B*), with the $t_2$ levels splitting further in vacancy complexes (*C*).

**Table Legends**

Table 1.   Defect level energies for various vacancy and NV centers in diamond and SiC (underlined

values indicate the level is occupied).

**Figure 1.**

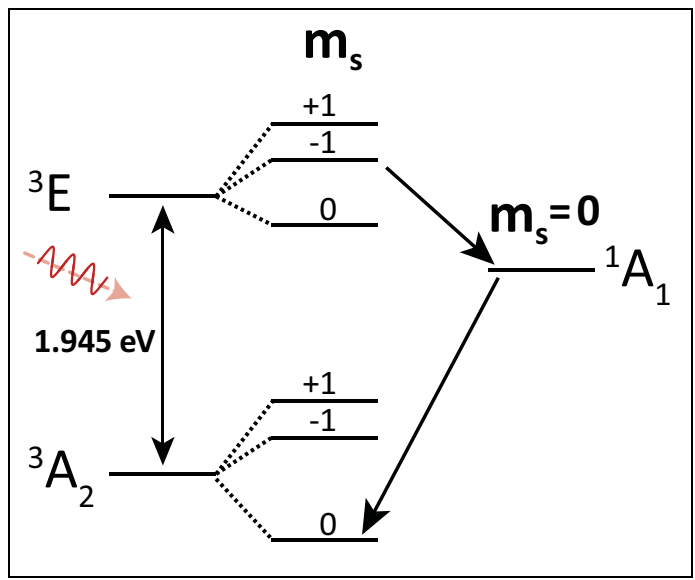

**Figure 2.**

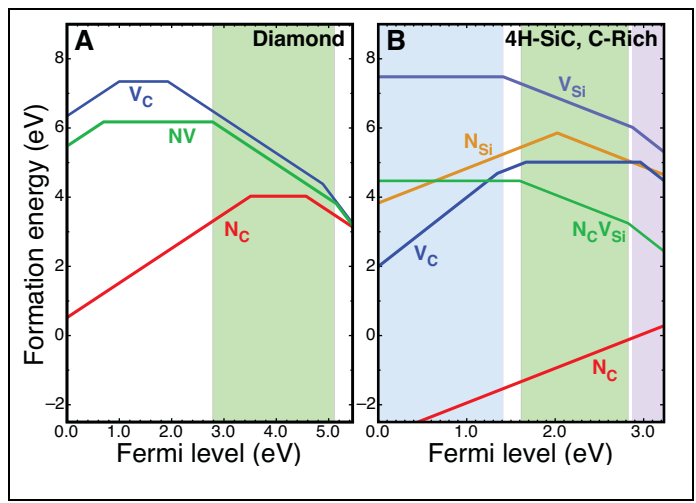

**Figure 3.**

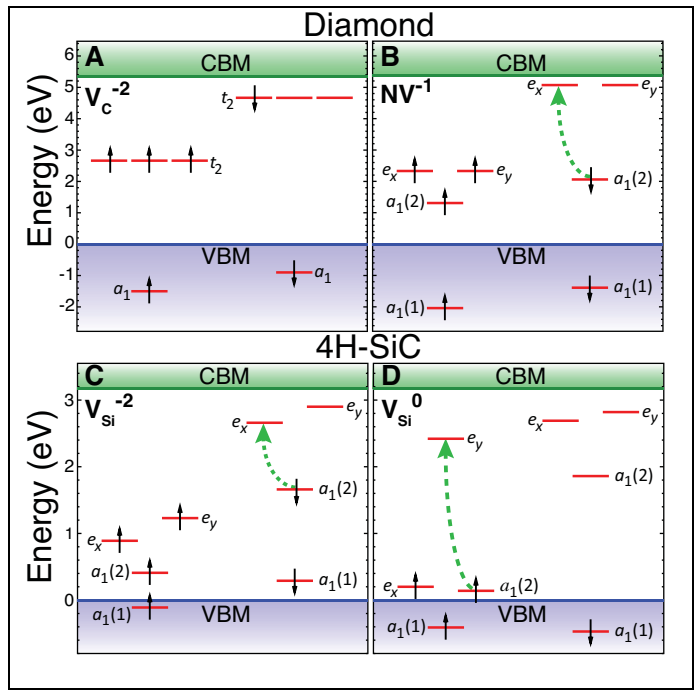

**Figure 4.**

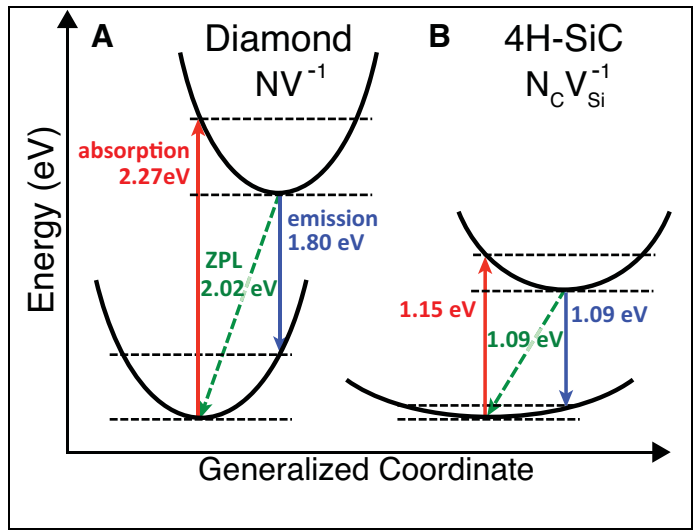

**Figure 5.**

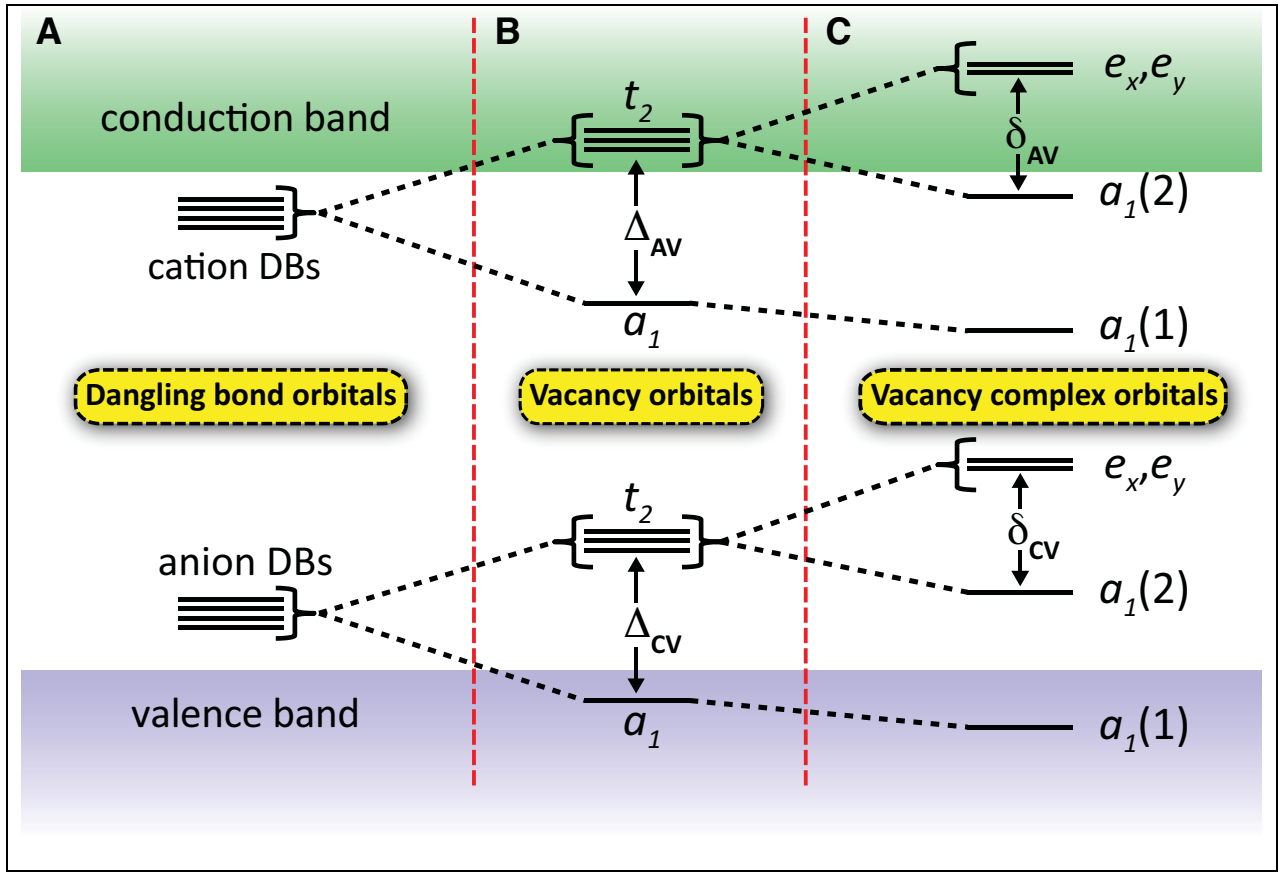

**Table 1.**

| Material | Defect | # of els. | spin majority | | | | spin minority | | | |
|---|---|---|---|---|---|---|---|---|---|---|
| | | | $a_1(1)$ | $a_1(2)$ | $e_x$ | $e_y$ | $a_1(1)$ | $a_1(2)$ | $e_x$ | $e_y$ |
| Diamond | $V^{-2}$ | 6 | <u>-1.46</u> | <u>2.66</u> | <u>2.66</u> | <u>2.66</u> | <u>-0.09</u> | <u>4.67</u> | 4.67 | 4.67 |
| Diamond | $NV^{-1}$ | 6 | <u>-2.04</u> | <u>1.31</u> | <u>2.33</u> | <u>2.33</u> | <u>-1.39</u> | <u>2.06</u> | 5.07 | 5.07 |
| 4H SiC | $N_C V_{Si}^{-1}$ | 6 | <u>-0.41</u> | <u>0.06</u> | <u>0.29</u> | <u>0.33</u> | <u>-0.72</u> | <u>0.93</u> | 2.35 | 2.51 |
| 4H SiC | $V_{Si}^{-2}$ | 6 | <u>-0.11</u> | <u>0.41</u> | <u>0.89</u> | <u>1.23</u> | <u>0.29</u> | <u>1.66</u> | 2.66 | 2.90 |
| 4H SiC | $V_{Si}^{-1}$ | 5 | <u>-0.11</u> | <u>0.43</u> | <u>0.43</u> | <u>0.67</u> | <u>0.53</u> | 2.66 | 2.67 | 3.14 |
| 4H SiC | $V_{Si}^{0}$ | 4 | <u>-0.41</u> | <u>0.14</u> | <u>0.20</u> | 2.42 | <u>-0.47</u> | 1.86 | 2.69 | 2.82 |

# Quantum computing with defects: Supporting Information

*J. R. Weber[1], W. F. Koehl[1], J. B. Varley[1], A. Janotti, B. B. Buckley,*
*C. G. Van de Walle, and D. D. Awschalom[2]*

*Center for Spintronics and Quantum Computation, University of California, Santa Barbara, CA 93106*

[1] *J.R.W., W.F.K., and J.B.V. contributed equally to this work.*
[2] *To whom correspondence should be addressed.  E-mail: awsch@physics.ucsb.edu*

### *Section S1: Identifying candidate host materials*

It is straightforward to identify tetrahedrally-coordinated semiconductors that satisfy criteria H1-H4 outlined in the main text.  First, band-structure parameters can be used to identify hosts that satisfy H1 and H2.  In Table S1, we list a number of tetrahedrally-coordinated hosts whose band-gaps are larger 2.0 eV (diamond, Si, and GaAs are listed in the bottom three rows for comparison.)  While 2.0 eV is an arbitrarily-defined value, we would like to accommodate deep centers whose optical transitions lie in the near-infrared ($0.89 - 1.65$ eV) or visible ($1.65 - 3.10$ eV) regions of the electromagnetic spectrum because optical equipment compatible with these energies is widely available.  The band-gap energy ($E_g$) of each host is listed in the second column.  In the third column, we list the spin-orbit splitting ($\Delta_{SO}$) of each material, as taken from valence-band splitting(s) at the $\Gamma$ point.  While $\Delta_{SO}$ is not a direct measure of the spin-orbit coupling in a host, it is still indicative of the strength of the spin-orbit interaction.  Values of $E_g$ and $\Delta_{SO}$ are room-temperature values unless otherwise noted.  In materials where more than one crystal structure is stable at room temperature, we have chosen to display the band parameters for the dominant room-temperature phase.

In the fourth column of the table, we list whether stable isotopes with nuclear spin equal to zero exist for the atomic species of each compound (criterion H4).  We note that while the lack of a nuclear spin bath may help to increase the spin-coherence time of a paramagnetic deep center, H4 is not necessarily a strict requirement.  It is beyond the scope of this paper to address the question of whether current growth technologies for each material are compatible with isotopic engineering, but it should be noted that the natural abundance of spin-0 isotopes varies by atomic species.

All of the hosts listed can be grown as single crystals, but the quality currently varies widely by material.  For instance, the types and numbers of extended defects that one may expect in a state-of-the-art growth of each host varies widely by material, as does the current maximum single-crystal size.  Nevertheless, many of the materials listed (such as 4H-SiC, ZnO, and GaN) can be bought commercially as wafers an inch or more in diameter.

### *Section S2*: *Trends in defect level splitting*

Although detailed calculations are necessary to systematically determine the splitting and location of defect levels, important insights are provided by the behavior of interacting DB orbitals that give rise to the defect levels, as shown in Fig. 5 of the main text.  These DB orbitals are closely related to the $sp^3$ orbitals in a tetrahedrally-coordinated semiconductor.  To demonstrate these concepts with a specific example, let us consider a cation vacancy (CV) in a tetrahedral semiconductor (surrounded by interacting anion DBs as depicted in the lower half of Fig. 5).  As discussed in the main text, the $t_2$ vacancy levels tend to be located in the lower half of the band gap.  Here we address how the choice of host and defect center impacts the energy position of the anion DB orbitals, and the splitting between the $a_1$ and $t_2$ vacancy orbitals ($\Delta_{CV}$ - Fig. 5$B$).

As the anion becomes more electronegative, i.e., closer to the upper right corner of the periodic table, the energy of its atomic $s$ and $p$ orbitals decreases and the orbitals become more localized. The overlap of these $sp^3$ DB orbitals determines the splitting $\Delta_{CV}$ between the vacancy levels as illustrated in Fig. 5B. This overlap is determined by the degree of localization of the $sp^3$ orbitals and by the spatial separation between the anions (Fig. S1A,B). If a substitutional impurity (X) is placed onto an A-site that neighbors the vacancy (Fig. S1C), the defect levels will further split into $a_1(1)$, $a_1(2)$, $e_x$, and $e_y$ as a consequence of the reduced symmetry.

These resulting defect levels can be filled with electrons in various ways, depending on the charge state. For defects analogous to the NV$^{-1}$ defect in diamond, the $a_1(1)$ level will be well below the VBM, as shown schematically in Fig. 2. Assuming this is the case, we can fill the remaining gap levels to obtain the desired spin-one configurations. Figure S2 outlines the two possibilities. Six electrons are needed to create a configuration similar to the NV$^{-1}$ in diamond, which has two electrons in the $a_1(2)$ level, and one electron in each of the $e$ levels (Fig. 2A). This allows for a spin-conserving transition between the spin-minority $a_1(2)$ and $e$ levels. In addition, a four-electron configuration also exists with a spin-triplet ground state (Fig. S2B), as was discussed for the $V_{Si}^0$ defect in SiC. This configuration allows for a spin-conserving transition between the spin-majority $a_1(2)$ and $e_y$ levels, as shown in Fig. 3D of the main text.

Now that we understand how many electrons are needed to form ground-state triplets in these configurations, we can determine which defect charge states are needed to produce such occupations. The charge state $Q$ of the defect is given by:

$$Q = 4 \times \frac{N_A}{4} + \left(N_X - N_A\right) - n_e = N_X - n_e \quad , \quad [S1]$$

where $n_e$ is the total number of electrons in the defect levels, and $N_A$ ($N_X$) the number of valence electrons for atom A (X). For example, $n_e = 6$ for the N$_C$V$_{Si}$ center in SiC, as discussed in the main text. Furthermore, $N_X = 5$ because atom X is a nitrogen atom. Hence, $Q = -1$, as noted in the main text.

As another example, consider the Zn vacancy (V$_{Zn}$) in ZnSe, which has been calculated to be stable in the 0, -1, and -2 charge states (4). In the neutral charge state ($Q=0$) of V$_{Zn}$, and with $N_X = N_A = 6$ (since there is no impurity present), Equation S1 shows that $n_e = 6$, and thus a spin-triplet ground-state similar to Fig. S2A is stable. In addition, we can place an impurity next to the vacancy. If we focus on Group-VII atoms, which act as donors on the oxygen site and are electrostatically attracted to the vacancy, $N_X = 7$ and $n_e = 6$, so that $Q = +1$.

In this discussion we have only considered the number of electrons needed to fill defect levels similar to those shown in Fig. S2. To properly address broader issues, such as whether defect levels are sufficiently deep in the band gap and whether criteria D1-D5 can be satisfied, explicit first-principles calculations are needed for each defect in question.

## References

1) Sridhara SG, Bai S, Shigiltchoff O, Devaty RP, Choyke WJ (2000) Differential absorption measurement of valence band splittings in 4H SiC. *Mater. Sci. Forum* 338–342:567-570.
2) Wei S-H, Zunger A (1996) Valence band splittings and band offsets of AlN, GaN, and InN. *Appl. Phys. Lett.* 69:2719-2721.
3) Lawaetz P (1971) Valence-band parameters in cubic semiconductors. *Phys. Rev. B* 4:3460–3467.
4) Laks DB, Van de Walle CG, Neumark GF, Blöchl PE, Pantelides ST (1992) Native defects and self-compensation in ZnSe. *Phys. Rev. B.* 45:10965.

## Figure Legends

Figure S1. Development of defect level structure in tetrahedrally coordinated (A-B) compound semiconductors. Atomic $sp^3$ dangling bonds ($A$) interact to form $a_1$ and $t_2$ levels in an ideal vacancy ($B$), with the $t_2$ levels splitting further in vacancy complexes ($C$).

Figure S2. Schematic defect-level diagrams for vacancy-related complexes in tetrahedrally coordinated semiconductors. These diagrams show defect levels with an occupation of six ($A$) and four ($B$) electrons.

## Table Legends

Table S1. All values of $E_g$ and $\Delta_{SO}$ are room-temperature values and are taken from Ref. 29 in the main text, unless noted otherwise. *Data taken from Ref. 1 of the Supporting Information. †Data taken from Ref. 2 of the Supporting Information. ‡Data taken from Ref. 3 of the Supporting Information.

**Figure S1.**

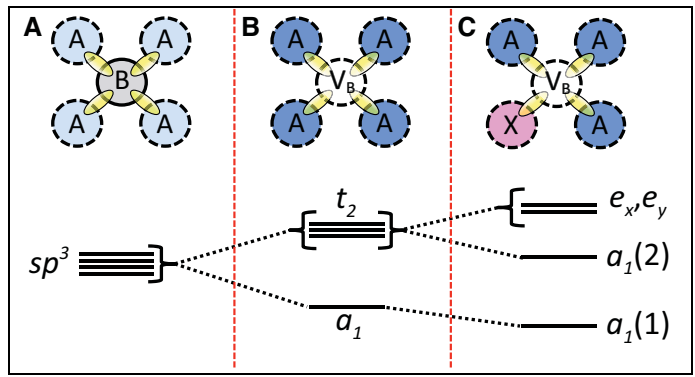

**Figure S2.**

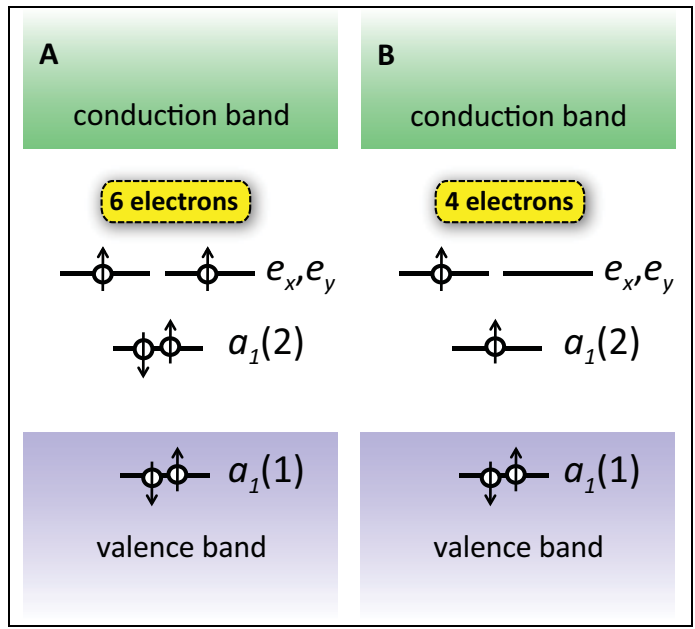

**Table S1. Host material parameters**

| Material | Bandgap, $E_g$ (eV) | Spin-Orbit Splitting, $\Delta_{SO}$ (meV) | Stable Spinless Nuclear Isotopes? |
|---|---|---|---|
| 3C-SiC | 2.39   (2 K) | 10   (2 K) | Yes |
| 4H-SiC | 3.26   (4 K) | 6.8   (2 K)[*] | Yes |
| 6H-SiC | 3.02   (4 K) | 7.1   (2 K) | Yes |
| AlN | 6.13 | 19 (Theory)[†] | No |
| GaN | 3.44 | 17.0   (10 K) | No |
| AlP | 2.45 | 50 (Theory)[‡] | No |
| GaP | 2.27 | 80 | No |
| AlAs | 2.15 | 275 | No |
| ZnO | 3.44   (6 K) | -3.5   (6 K) | Yes |
| ZnS | 3.72 | 64 | Yes |
| ZnSe | 2.82   (6 K) | 420 | Yes |
| ZnTe | 2.25 | 970   (80 K) | Yes |
| CdS | 2.48 | 67   (10 K) | Yes |
| Diamond (C) | 5.5 | 6   (1.2 K) | Yes |
| Si | 1.12 | 44   (1.8 K) | Yes |
| GaAs | 1.42 | 346   (1.7 K) | No |